\documentclass[aps,prb,twocolumn,showpacs,superscriptaddress]{revtex4-1}
\usepackage{graphicx}

\begin{document}

\title{Antiferromagnetic Critical Fluctuations in BaFe$_2$As$_2$}

\author{Stephen D. Wilson}
\email{stephen.wilson@bc.edu}
\affiliation{ Materials Science Division, Lawrence Berkeley National Lab, Berkeley, Caifornia 94720, USA }
\affiliation{ Department of Physics, Boston College, Chestnut Hill, Massachusetts 02467, USA }
\author{Z. Yamani}
\affiliation{ Chalk River Laboratories, Canadian Neutron Beam Centre, National Research Council, Chalk River, Ontario, Canada K0J 1P0} 
\author{C. R. Rotundu}
\affiliation{ Materials Science Division, Lawrence Berkeley National Lab, Berkeley, Caifornia 94720, USA}
\author{B. Freelon}
\affiliation{ Physics Department, University of California, Berkeley, California 94720, USA}
\author{P. N. Valdivia}
\affiliation{ Materials Science Department, University of California, Berkeley, California 94720, USA}
\author{E. Bourret-Courchesne}
\affiliation{ Materials Science Division, Lawrence Berkeley National Lab, Berkeley, Caifornia 94720, USA }
\author{J. W. Lynn}
\affiliation{ NIST Center for Neutron Research, National Institute
of Standards and Technology, Gaithersburg, Maryland 20899, USA}
\author{Songxue Chi}
\affiliation{ NIST Center for Neutron Research, National Institute
of Standards and Technology, Gaithersburg, Maryland 20899, USA}
\affiliation{Department of Materials Science and Engineering, University of Maryland, College Park, Maryland 20742, USA}
\author{Tao Hong}
\affiliation{ Neutron Scattering Science Division, Oak Ridge National Laboratory, Oak Ridge, Tennessee 37831-6393, USA}
\author{R. J. Birgeneau}
\affiliation{ Materials Science Division, Lawrence Berkeley National Lab, Berkeley, Caifornia 94720, USA }
\affiliation{ Physics Department, University of California, Berkeley, California 94720, USA}
\affiliation{ Materials Science Department, University of California, Berkeley, California 94720, USA}
\begin{abstract}

Magnetic correlations near the magneto-structural phase transition in the bilayer iron pnictide parent compound, BaFe$_2$As$_2$, are measured.  In close proximity to the antiferromagnetic phase transition in BaFe$_2$As$_2$, a crossover to three dimensional critical behavior is anticipated and has been preliminarily observed.  Here we report complementary measurements of two-dimensional magnetic fluctuations over a broad temperature range about T$_N$.  The potential role of two-dimensional critical fluctuations in the magnetic phase behavior of BaFe$_2$As$_2$ and their evolution near the anticipated crossover to three dimensional critical behavior and long-range order are discussed.   
\end{abstract}

\pacs{74.70.Dd, 75.25.z, 75.50.Ee, 75.40.Cx}

\maketitle

\section{Introduction}
Reduced dimensionality constitutes a key element of the magnetism inherent in the lamellar copper oxides and is also thought to play a key role within the mechanism responsible for the formation of the high temperature superconducting (high-T$_c$) phase \cite{kastner}.  The recent discovery of superconductivity at temperatures as high as T$_{c}=55$K within an alternate class of iron based superconductors\cite{kamihara, rotter, chen, rotter2} has stimulated a host of comparisons between the ferrous parent compounds of these new superconductors and the Mott-insulating parent systems of the cuprates.  In particular, antiferromagnetic order exhibiting energetic and anisotropic magnetic exchange interactions is observed within both classes of materials\cite{kastner, coldea, zhaoCaFe2As2,dialloCeFe2As2}.  While the spin ordering in high-T$_c$ cuprates such as La$_{2}$CuO$_4$ has been demonstrated to arise from strong two-dimensional (2D) spin fluctuations consistent with a 2D-Heisenberg model with weak interlayer coupling\cite{birgeneauLa2CuO4, coldea}, the buildup of correlations yielding the antiferromagnetic parent phase in the iron based superconductors remains relatively unexplored.  

Specifically, within the bilayer iron pnictide variant BaFe$_2$As$_2$ (Ba-122) studies of both pure and doped samples have suggested a second order or weakly first order phase transition at T$_N=136$K \cite{wilsonBaFe2As2, matan, johrendt, mcqueeny, lelandnidoped}.  Critical fluctuations in proximity to the onset of long-range magnetic order have long been known to yield important insight into the fundamental spin and spatial symmetries driving the phase formation.  It is therefore important to explore the nature of the fluctuations driving the magnetic phase transition in Ba-122 as a means of understanding the spin behavior in this iron pnictide parent system as it passes through its simultaneous magnetic and structural phase transitions.  

BaFe$_2$As$_2$ develops long range antiferromagnetic order at T$=136$K concomitant to a structural distortion from tetragonal to orthorhombic symmetry\cite{huangBaFe2As2}.  The biquadratic free energy coupling term between the structural and magnetic order parameters (implied by their identical phase behaviors) suggests an unorthodox coupling scheme rather than a simple linear coupling of the magnetic order parameter squared to the lattice through strain\cite{wilsonBaFe2As2, jesche, lester}.  Models such as fluctuating magnetic domains locally breaking time reversal symmetry prior to the transition to long-range order \cite{mazin} or the supposition of a weakly first order nature to the structural phase transition may explain this peculiar coupling term; however other symmetries such as higher energy orbital ordering\cite{chenorbital, leeorbital} or valley density waves \cite{tesanovic} must be considered as well.  

In this article, we present a neutron scattering study of 2D magnetic fluctuations above and below the antiferromagnetic phase transition in BaFe$_2$As$_2$.  Given the 2D-Ising-like exponent of the magnetic phase transition\cite{wilsonBaFe2As2, wilsonuniversality}, 2D-Heisenberg fluctuations are expected at temperatures far above T$_N$.  As the system is cooled sufficiently close to $T_N$, the spin system must crossover to three-dimensional behavior, and the longitudinal component of these two-dimensional fluctuations is expected to condense into the three-dimensional Bragg positions for the antiferromagnetic ordered phase.  Transverse spin fluctuations however are expected to remain noncritical and simply evolve continuously into the spin wave component of the spectrum as the system is tuned across T$_N$.  Here we present new measurements of the energy-integrated, two dimensional fluctuations in Ba-122 that are consistent with this picture, and these results complement our previous observation of 3D critical fluctuations persisting until T$\approx 150$K ($1-\frac{T}{T_N}\equiv$t$\approx 0.10$) in the same sample\cite{wilsonBaFe2As2}.  Our study supports the notion that the magnetic phase transition in the BaFe$_2$As$_2$ system is driven fundamentally by magnetic fluctuations of two dimensional character.      

\begin{figure}
\includegraphics[scale=.4]{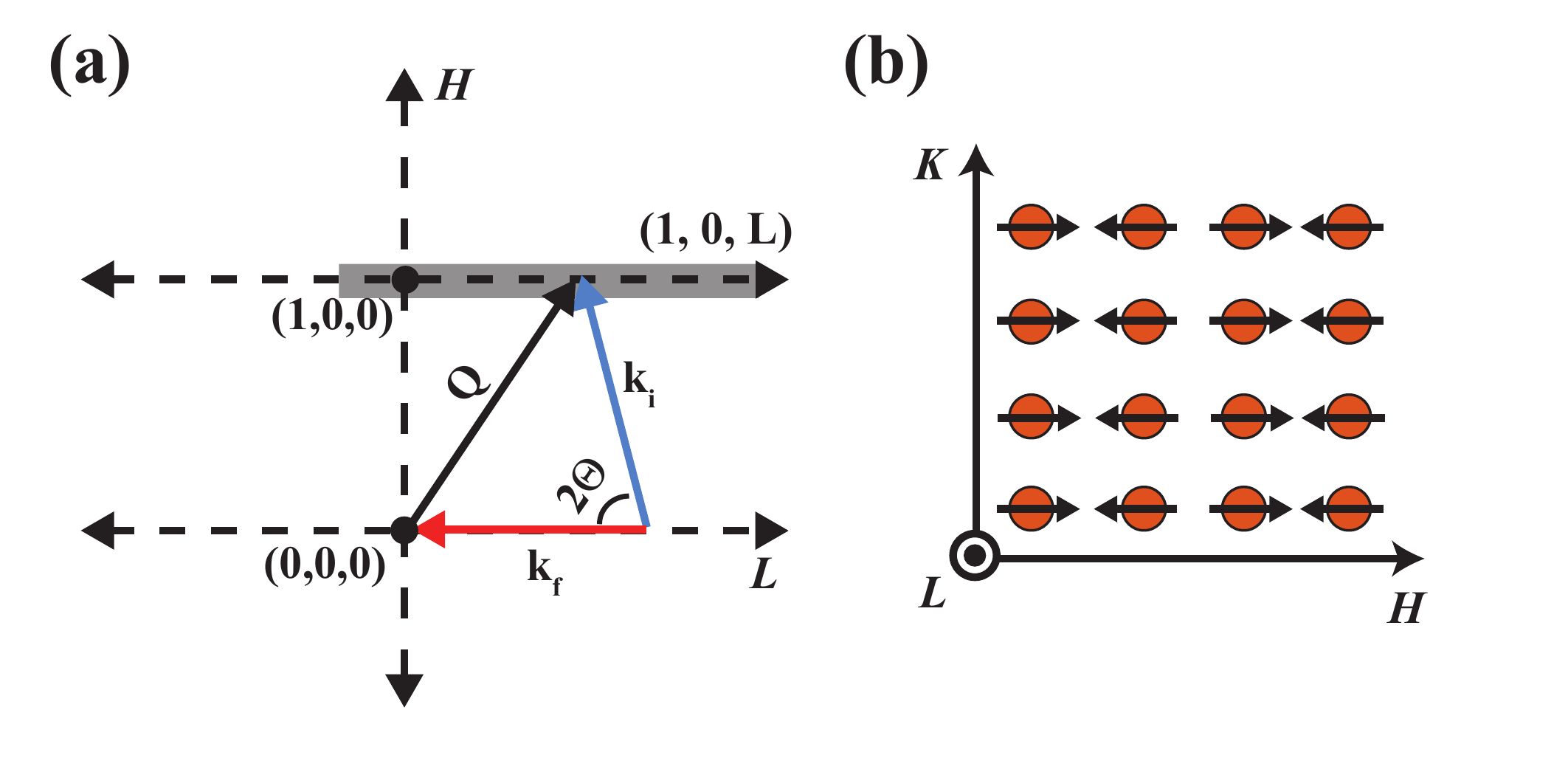}
\caption{(a) Schematic showing the scattering configuration employed to measure the energy integrated, instantaneous correlation length with $c\parallel k_f$ (b) Two-dimensional spin structure for BaFe$_2$As$_2$ showing spin orientations within the Fe-plane only.  Critical fluctuations measured within the spectrometer configuration in panel (a) at non-integer L necessarily arise from uncorrelated planes with this in-plane ordering pattern.}
\end{figure}

\section{Experimental} 
A $600$mg crystal of BaFe$_2$As$_2$ was grown out of an FeAs flux with a ratio Ba:Fe:As of 1:4:4 and possessed a mosaicity of less than 0.3$^\circ$ FWHM.  The crystal was mounted in the [H, 0, L] plane (orthorhombic notation $a=b=5.6202\AA$, $c=13.0323\AA$ at T$=300$K) inside of a closed cycle refrigerator. Positions in reciprocal space are given in reduced lattice notation where \textbf{$\vec{Q}$} [r.l.u.] $=$\textbf{$\vec{Q}$}[$\AA^{-1}$]$\cdot(\frac{a}{2\pi}\hat{h}+\frac{b}{2\pi}\hat{k}+\frac{c}{2\pi}\hat{l})$. Experiments were performed on the BT-7 triple-axis spectrometer at the NIST Center for Neutron Research at the National Institute of Standards and Technology and on the HB-1 triple-axis spectrometer at the High Flux Isotope Reactor at Oak Ridge National Lab.  Experiments on BT-7 were performed with an incident neutron energy of E$_i=14.7$meV and collimations of open-50$^{\prime}$-sample-50$^{\prime}$-detector.  Experiments on HB-1 were performed with an incident energy of E$_i=13.5$meV and collimations of 40$^{\prime}$-40$^{\prime}$-sample-60$^{\prime}$-240$^{\prime}$-detector.  In all experiments, two PG filters were placed before the sample in order to eliminate higher order contamination.  The (0, 0, 2) reflections of pyrolitic graphite (PG) crystals were used as the incident energy on vertically focusing monochromaters. Uncertainties presented in the data are statistical in origin and represent one standard deviation.  Experiments were performed with the analyzer crystal removed so that two-axis energy integration along the k$_f$ side could be performed.   

\section{Magnetic Scattering in B\MakeLowercase{a}F\MakeLowercase{e}$_2$A\MakeLowercase{s}$_2$}
In order to isolate the inherently 2D component of the spin scattering in Ba-122 resulting from magnetism correlated only within the Fe-planes, the experimental geometry was chosen such that the c-axis was parallel to the final scattering wave vector, k$_f$\cite{birgeneauK2NiF4, alsnielsen}.  The analyzer was removed thus ensuring that any two-dimensional diffuse scattering along L at finite energy transfers maintains the specified in-plane H-value (Fig. 1).  When integrating over various energy transfers at a fixed two-theta value, the corresponding momentum transfers probed by the neutron must also vary; however through orienting $c\parallel$k$_f$ the component of momentum transfer that varies with final energy is constrained to be along only the L-direction in momentum space.  For scattering that is diffuse along L, this technique is uniquely powerful in providing an efficient tool for the measurement of an energy integrated spin response.  This allows the instantaneous correlation function for any 2D component of magnetic scattering within the Fe-planes to be measured through scans along H with fixed k$_i$-dependent L-values.  Furthermore, our geometry is also an efficient method for probing the presence of diffuse, quasi-2D scattering through searching for magnetic signal at nominally disallowed 3D magnetic Bragg positions.  For instance, when probing the in-plane magnetic zone center at H$=1$ with E$_i=14.7$ meV, this translates to a geometry where the c-axis is parallel to k$_f$ at an L-value of L$=0.51$.  

Energy transfers probed by neutron energy loss are integrated up E$_i$ while all energy transfers probed by neutron energy gain are integrated.  The most relevent range of integrated L values, assuming neutron energy transfers in the window from $-11.7\leq\Delta E\leq14.7$ meV, can be estimated as $-5.01\leq$L$_{int}\leq2.39$ (r.l.u.).  This range accounts for the most significant thermally activated fluctuations at T$_N$.  Thus, the presence of correlated magnetic scattering at the (1, 0, 0.51) position in this geometry provides qualitative evidence of 2D diffuse scattering integrated within the detector rather than dispersive 3D excitations that would cut through only a small fraction of our experimental window and be lost in the larger background incurred via this technique.       

In Fig. 2 (a), we have plotted the raw data for scans along the [H, 0, 0] direction with L=0.51 (this satisfies c$\parallel k_f$ for E$_i=14.7$meV).  Data taken at T=T$_N$ are overplotted with data collected far below the transition temperature at T=$80$K.  From these data, it is immediately evident that at the transition a substantial amount of scattering appears centered at the Q$_{in-plane}=(1, 0)$ position.  Subsequent H-scans for T$<100$K show no further changes in intensity at Q$_{in-plane}=(1, 0)$ upon cooling, indicating that the 2D fluctuations have disappeared by this temperature. Both H- and L-scans below these temperatures
(blue symbols) therefore serve as a useful background for the removal of nonmagnetic scattering and for the isolation of critical scattering near $T_N$.  Similarly, the small peak at H$=1.25$ in Fig. 2 (a) is a temperature independent nuclear peak from another small crystallite with the sample. The background subtracted data at T$=136$K in Fig. 2 (c) show a clear peak centered at the Q$_{in-plane}=(1, 0)$ position indicative of scattering arising from quasi-elastic 2D, in-plane fluctuations at the AF transition.  In order to prove that the signal observed at Q$_{in-plane}=(1, 0)$ is indeed two-dimensional, scans were performed along L, and plotted in Figs. 2 (b) and (d) showing corresponding raw and subtracted L-scans respectively.  The subtracted scan in Fig. 2 (d) shows the critical scattering peaks at the two-axis integration condition of $c\parallel k_f$ at L$=0.51$---as expected for 2D diffuse scattering at the Q$_{in-plane}=(1, 0)$ position.  This broad peak, extending over half of the Brillouin zone in L, sharply contrasts with the relatively sharp peak observed in H-scans and is indicative of 2D fluctuations arising from spins associated with the antiferromagnetic phase. However, it is important to note that the width of the peak in L is predominantly controlled by the relative orientation of c with respect to k$_f$ (ie. the degree of integration performed rather than the correlations between the planes).  The intrinsic spectrometer resolution is a neglible component of the L-scan's peak width as shown in Fig. 2. 

\begin{figure}[t]
\includegraphics[scale=.4]{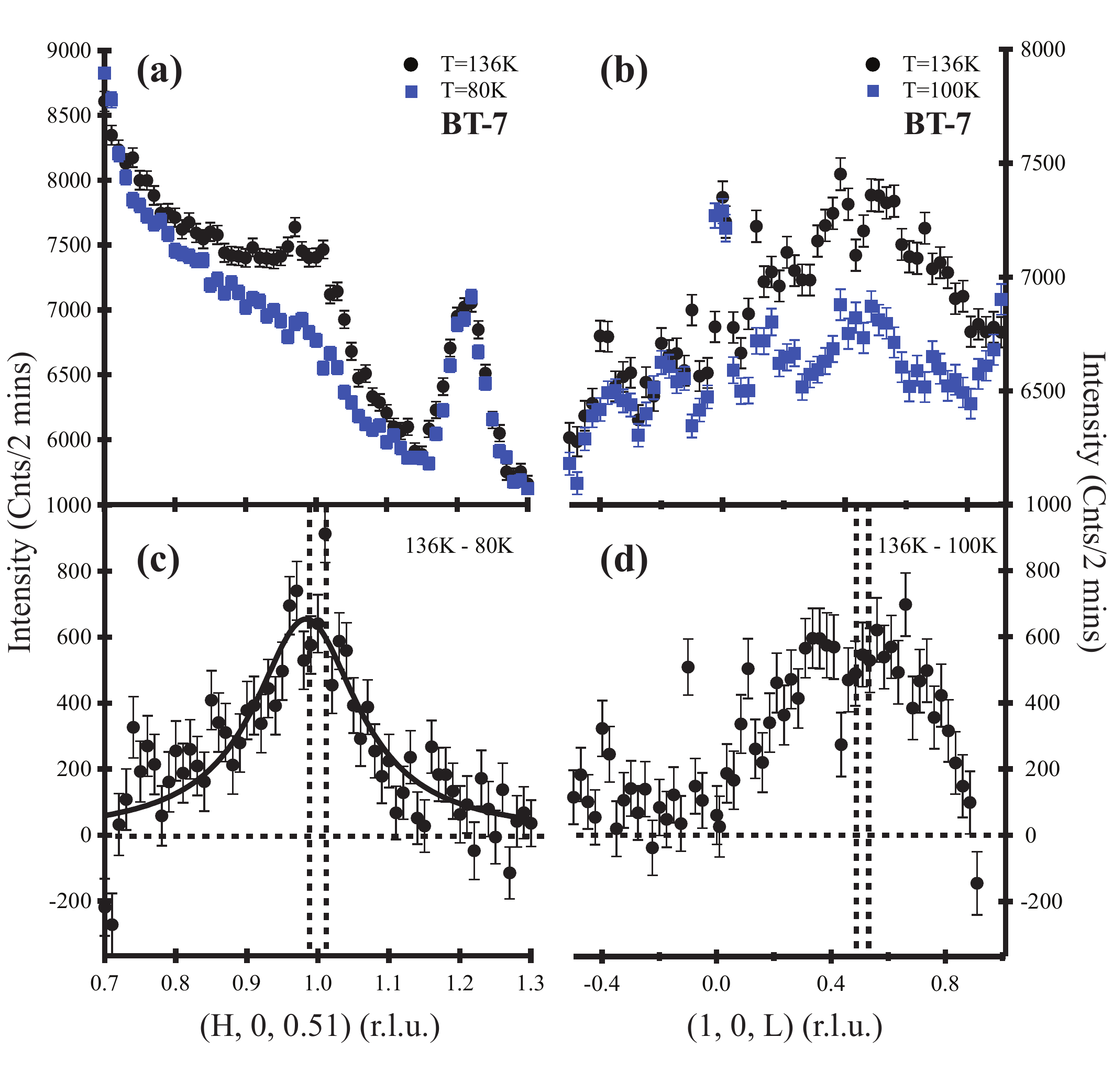}
\caption{(a) H-scans through the energy integration position (1, 0, 0.51) both below (blue square symbols) and at T$_N$ (black circle symbols).  (b)  Raw data from L-scans through (1, 0, 0.51) below and at T$_N$ respectively. Background-subtracted data for both H-scans and L-scans through (1, 0, 0.51) are also plotted in panels (c) and (d) respectively.  Vertical dashed lines in panels (c) and (d) illustrate the full width at half maximum (FWHM) of the spectrometer resolution function projected along the H and L directions respectively.}
\end{figure}

2D short-range spin correlations persist far above T$_N$ at the AF ordering wave vector and eventually vanish by $\approx250$K.  Figure 3 (a) shows background-subtracted H-scans through the $(1,0)$ position at $136$K and $300$K, revealing the absence of the 2D scattering signal at high temperatures.  A gradual increase in the overall background intensity was observed above $150$K and is reflected by the large uniform offset in the $300$K subtracted data.  

In investigating the detailed temperature dependence of the critical scattering signal, the peak intensity of the scattering at $(1, 0)$ was measured as the sample was slowly warmed through T$_N$.  The resulting intensity as a function of temperature is plotted in Fig. 3 (b) where the intensity of the 2D scattering sharply increases upon approaching T$_N$ from below and gradually decreases as the system is warmed away from the transition temperature.  In order to compensate for the increase in background above $150$K, the nonmagnetic background contribution to the scattering was measured by rotating the crystal $15^{\circ}$ away from the (1, 0)$_{in-plane}$ position, and it was subsequently removed from the data shown in Fig. 3 (b). By $250$K the peak has broadened substantially and is beyond the experiment's ability to resolve it above the background. 

The intensity of the 2D spin scattering peaks in close proximity to $136$K near the onset of 3D AF order and is consistent with the scattering intensities at finite energy transfer reported in CaFe$_2$As$_2$ (Ca-122)\cite{diallo}.  To provide a better reference for this, the purely elastic ($\Delta$E$=0$) intensity of the Q$=(1, 0, 3)$ position is overplotted with the 2D scattering intensity in Fig. 3 (b).  The sharp decrease in 2D scattering intensity below T$_N$ likely reflects the spin gap opening below T$_N$ and the subsequently reduced scattering intensity as it continuously approaches its full $\Delta=10$meV value\cite{matan}. The buildup in scattering intensity in approaching T$_N$ from above reflects the buildup of in-plane correlations, both longitudinal and transverse components, as the system is cooled toward the phase transition.  The dominant contribution to this signal should originate from the transverse, spin wave, component of the 2D scattering, and this will be discussed in more detail in the Discussion section of this paper.  

\begin{figure}[t]
\includegraphics[scale=.4]{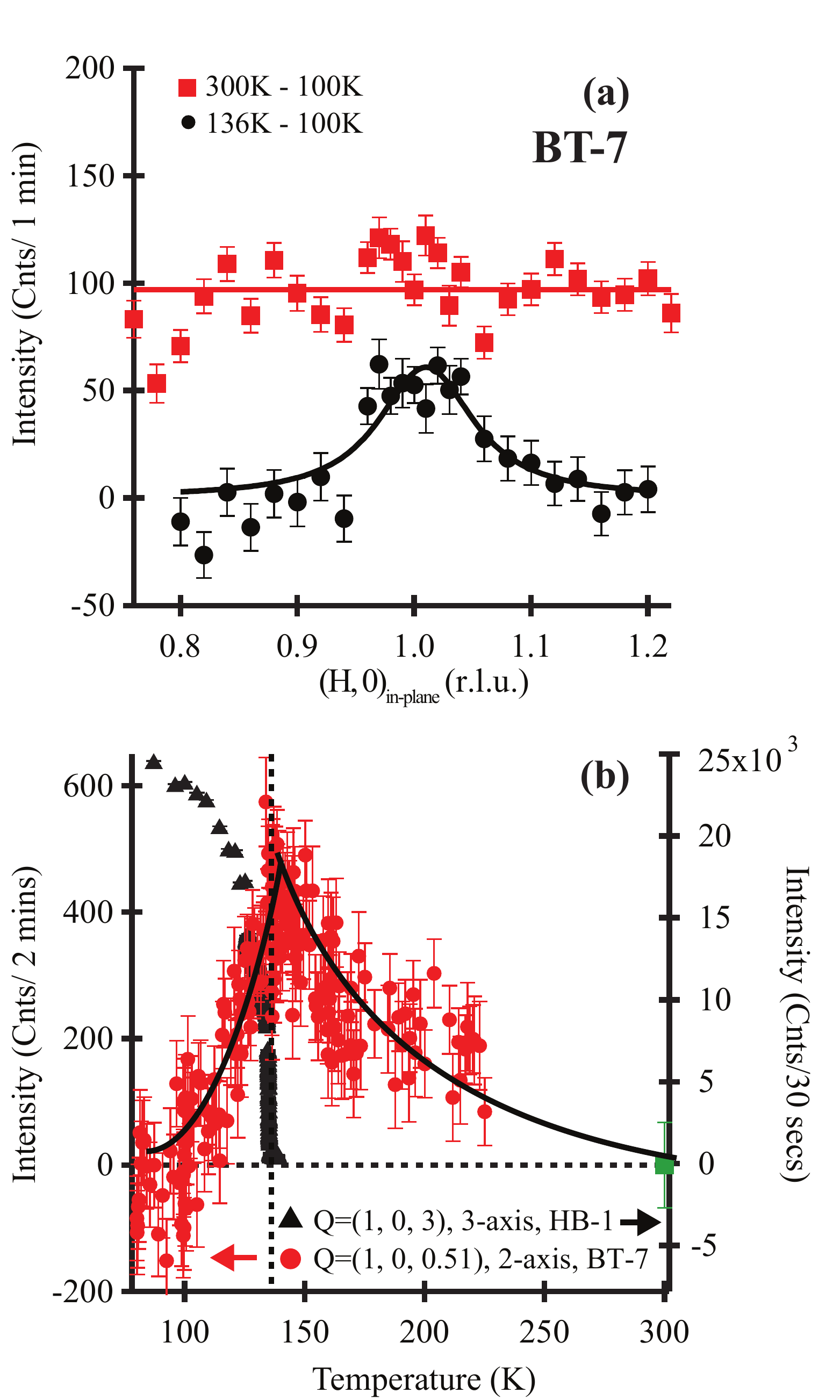}
\caption{Critical scattering signal in BaFe$_2$As$_2$.  (a) H-scans showing background subtracted data at 136K (black circles) and 300K (red squares) collected on BT-7.  (b) BT-7 data showing the background subtracted energy integrated intensities of 2D fluctuations plotted as a function of temperature (red circles).  The uncertainty for the green circle shows the uncertainty in a vanishing signal at 300K.  The vertical dashed line denotes T$_N$, and the solid black line through the data is a guide to the eye.  Elastic data collected at Q=(1, 0, 3) in a triple-axis configuration on HB-1 are also plotted as black triangles and show the development of the magnetic order parameter in the sample.}
\end{figure}

\begin{figure}[t]
\includegraphics[scale=.5]{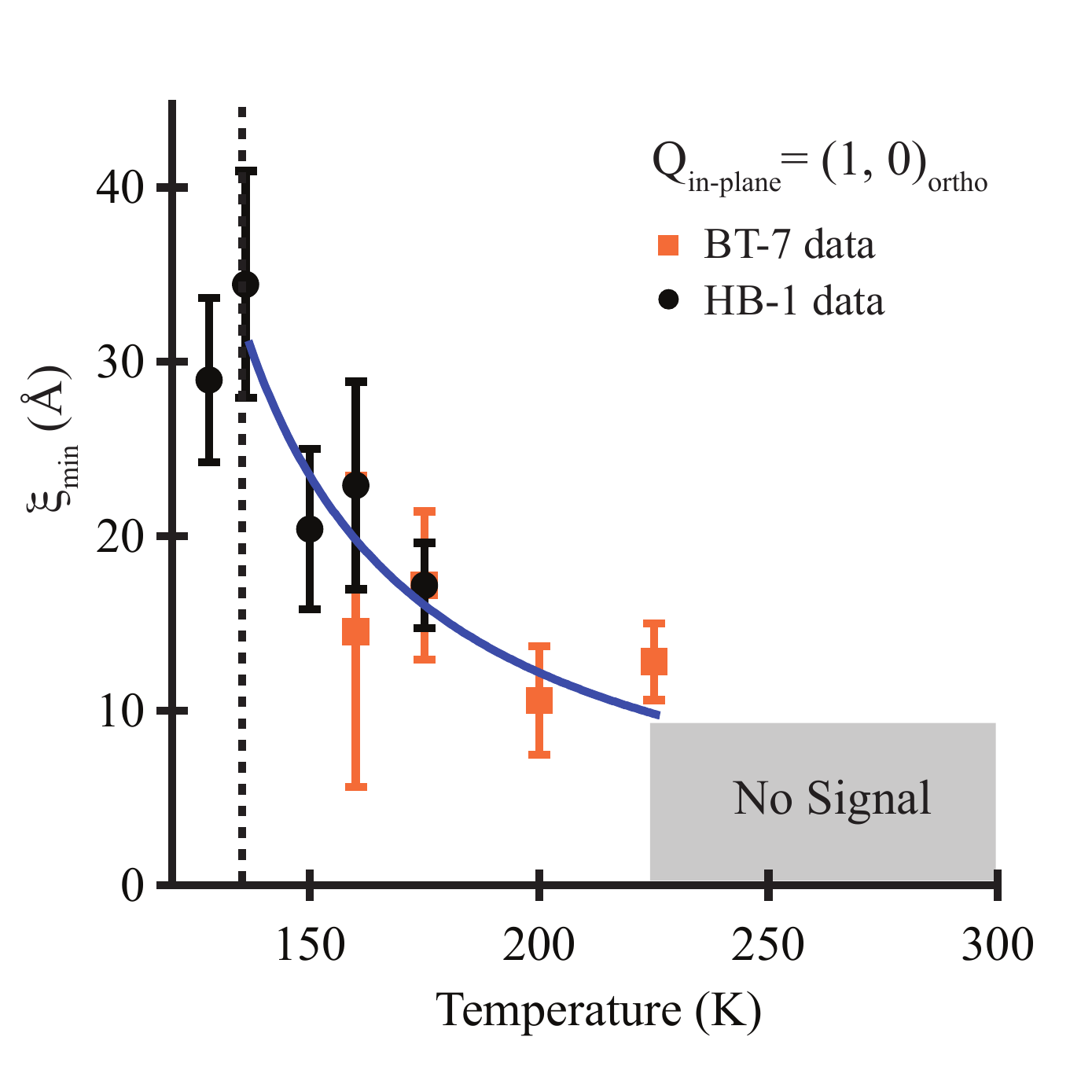}
\caption{2D, in-plane, spin correlation lengths measured in the two-axis configuration both above and below T$_N$.  The vertical dashed line denotes T$_N$.  Both E$_i=13.5$ and $14.7$ meV were used and scans conducted along H centered at the corresponding (1, 0, L) integration positions.  Data were collected on both the HB-1 and BT-7 spectrometers, and the solid line is a guide to the eye.}
\end{figure}

We next investigated the manner through which the correlation length of the 2D fluctuations evolved as the sample was cooled toward T$_N$.  In principle, this represents an approximation of the energy-integrated, instantaneous correlation function; however the energetic spin excitations in BaFe$_2$As$_2$ preclude a total integration.  Instead, only the most relevant, low-energy, fluctuations are probed with the incident energies used.  The background-subtracted data were fit to a two-dimensional Lorentzian profile convolved with the instrumental resolution. Within this convolution, the elastic resolution of the spectrometer was convolved into the fit peaks, which effectively underestimates the total resolution contribution to the scattering profile.  The differing resolution functions for dynamic correlations from finite energy transfers must be accounted for; however without knowing the precise spectral distribution of the local spin susceptibility for each corresponding temperature, this is not possible.  Instead, the correlation lengths plotted represent the minimum spin-spin correlation length at each temperature.  The results of fits at various temperatures are shown in Fig. 4.  At high temperatures far from T$_N$ the correlations are short-ranged, and as the system is cooled toward T$_N$ the correlation lengths begin to increase.  Correlation lengths smoothly increase upon decreasing temperature and seemingly saturate around T$_N$.  The absence of a true divergence to long-range order at the phase transition is expected given that the critical, longitudinal, component of the scattering necessary condenses into the 3D Bragg position as the system crosses over from 2D to 3D regimes above T$_N$. Hence, near T$_N$ we are measuring the noncritical, transverse correlation length.

\section{Discussion}
The two-dimensional critical scattering observed in BaFe$_2$As$_2$ far above T$_N$ is consistent with the 2D Heisenberg spin excitations anticipated at temperatures far above the 2D Ising-like magnetic order parameter in this system.  At these high temperatures, the scattering represents a mixture of both longitudinal and transverse 2D critical fluctuations.  Projecting the components of the measured susceptibility for an untwinned crystal into vectors perpendicular and parallel to the ordered moment direction ($\hat{\eta}=(1, 0, 0)$), the measured spin susceptibility can be expressed as the linear combination $\frac{\partial\sigma}{\partial\Omega}=\chi^{\perp}(1-sin^2(\theta))+\chi^{\parallel}(sin^2(\theta))$ where $\theta$ is the angle between the Q-vector and the $a$-axis and $\chi^{\perp}$ and $\chi^{\parallel}$ are projections of the spin susceptibility within the $(b,c)$-plane and along the $a$-axis respectively\cite{alsnielsen}. For a twinned crystal at the Q$=(1, 0, 0.51)$ position this becomes $\frac{\partial\sigma}{\partial\Omega}=1.477\chi^{\perp}+0.523\chi^{\parallel}$ where approximately $5\%$ of the measured signal is comprised of longitudinal fluctuations, $\chi^{\parallel}$. 

Upon cooling, the longitudinal component of the critical scattering is expected to condense into the 3D AF Bragg positions at the dimensional crossover T$_{3D}\approx150$K inferred from our previous studies\cite{WilsonBaFe2As2}.  The remaining transverse component represents the spin wave contribution to $\chi^{\prime\prime}$ which is nondivergent at T$_N$. This picture is consistent with the observed behavior of the 2D fluctuations in both Figs. 3 and 4. Due to the minority component of $\chi^{\parallel}$ in the measured 2D signal and the known crossover in the dimensionality of this system away from T$_N$, the correlation lengths plotted in Fig. 4 reflect the evolution of the spin-spin correlations of in-plane spin wave excitations. Below T$_N$ within the resolution of our experiments, we could detect no further change in the 2D correlation lengths; however a rapid decrease in scattering signal prevented measurements far below T$_N$. 

The buildup in low energy spectral weight upon approaching T$_N$ from above is consistent with the spin behavior observed within isostructural Ca-122 \cite{diallo} and with the expected divergence of the staggered spin susceptibility upon approaching T$_N$.   Upon cooling from high temperatures, 2D critical fluctuations build up within the longitudinal component $\chi^{\parallel}$ which are subsequently interrupted at T$_{3D}$, and our data suggest that the remaining transverse fluctuations continue to build smoothly toward T$_N$.  This conjecture is based on our expectation that the nondivergent behavior at T$_N$ is suggestive of remnant transverse fluctuations.  Below T$_N$, the intensity is therefore representative of only the transverse critical fluctuations that are subsequently diminished upon further cooling. The suppression of quasi-2D scattering upon cooling below T$_N$ is primarily due to the opening of a spin anisotropy gap within the ordered phase\cite{matan} whose value increases rapidly away from the 3D zone center with the buildup of interplane correlations below T$_N$ and the resulting dispersion of spin excitations along L.   

Another issue worth commenting on is the lack of any detectable anomaly in the quasi-2D signal in Fig. 3 at T$=150$K upon approaching T$_N$ from above.  Naively, one may expect to resolve some shift in the intensity of the 2D fluctuations as the longitudinal component of the signal condenses into the 3D AF ordering wave vectors near T$_{3D}$.  Dimensional crossovers however typically occur over the span of a decade in reduced temperature rendering changes resolvable primarily through shifts in the critical exponents of the phase transition.  Therefore, our current data preclude any definitive statement on the presence or absence of a crossover primarily due to insufficient statistics, although future experiments with increased sample volume will help resolve this issue.      

In attempting to analyze the data below T$_N$, one must consider the rapidly evolving dispersion as the system is cooled through both the magnetic and structural phase transitions at T$=136$K.  At present, there has been no detailed study of the evolution of magnetic exchange couplings where the nearest neighbor J$_{1A}$=J$_{1B}$ above T$_N$ transitions into a highly anisotropic antiferromagnetic J$_{1A}$ and ferromagnetic J$_{1B}$ below T$_N$\cite{zhaoCaFe2As2}.  Until the nature of this transition is known in detail, the complexity of the magnetic phase transition precludes a more in-depth analysis of the critical behavior below T$_N$.  The small sample volume of the Ba-122 crystal in the current study is the limiting factor rather than an intrinsically weak critical scattering signal\cite{birgeneauK2NiF4}, and future studies with a greatly increased sample volume will allow a more comprehensive evaluation of the critical behavior in this material.  

\section{Conclusions}
We have investigated the behavior of 2D spin fluctuations about the magnetic phase transition in the Ba-122 system.  From this initial study, we have resolved a mixture of both transverse and longitudinal critical fluctuations whose onset is coupled to appearance of 3D AF at T$_N$.  This suggests that 2D fluctuations play an important role in driving the formation of the magnetic phase within this system; however the interrupted divergence in the instantaneous correlation lengths and crossover to 3D spin behavior within this 2D magnetic scattering require further investigation.  Future experiments with larger sample volume and tighter experimental collimation will allow a more quantitative assessment of energy integrated, spin-spin correlation lengths close to T$_N$.

\acknowledgments{
This work was supported by the Director, Office of Science, Office of Basic Energy Sciences, U.S. Department of Energy under Contract No. DE-AC02-05CH11231 and Office of Basic Energy Sciences U.S. DOE under Contract No. DE-AC03-76SF008.  Work at ORNL was partially supported by the Division of Scientific User Facilities, Office of Basic Energy Sciences, US DOE.}


\end{document}